# Using a Metasurface to Enhance the Radiation Efficiency of Subterahertz Antennas Printed on Thick Substrates

Yevhen M. Yashchyshyn[1], Peter L. Tokarsky[1,2]

[1]Warsaw University of Technology, Institute of Radioelectronics and Multimedia Technology, ul. Nowowiejska 15/19, 00-665 Warsaw, Poland
[2]Institute of Radio Astronomy of the National Academy of Sciences of Ukraine,4, Mystetstv St., Kharkiv, 61002, Ukraine

**Abstract–**This study investigates the possibility of increasing the radiation efficiency of printed antennas and arrays by suppressing their inherent surface waves using a metasurface made of quad-split rings (QSR). A symmetrical resonant microstrip dipole and a four-element series-fed dipole array printed on an infinite grounded dielectric layer (layer thickness: 0.2 mm; relative permittivity: 9.4; tan delta: 0.0005) were simulated with FEKO 2022 software. Conducted at 100–116 GHz, the numerical results revealed extremely low radiation efficiencies of approximately 31% and 40% for the studied dipole and dipole array, respectively, which resulted from the presence of surface waves in the dielectric. However, placing only one QSR near each dipole arm triggered an increase in radiation efficiency by 2.5 times (up to 75%). The use of a metasurface in the form of two small QSR arrays triggered a pronounced improvement in radiation efficiency, reaching 93.6% and 96.5% for the studied dipole and dipole array, respectively. Analysis of the electric field distribution images showed that this enhancement resulted from surface wave suppression.

**Key words:** Printed antenna, radiation efficiency, thick substrate, metasurface unit cell, surface-wave suppression.

## Introduction

Printed antennas, known for their advantages such as lightweight, low-profile design, affordability, seamless integration with other devices, and efficient mass production, are extensively applied in the microwave frequency range. The likelihood of their widespread adoption extends to the subterahertz (Sub-THz) frequency range (100–300 GHz). However, most printed antennas exhibit a narrow relative bandwidth (2%–3%), which often limits their capabilities in the microwave range. Concurrently, at sub-THz, the 2%–3% band is 2–6 GHz, which allows devices to realize the benefits of this band, such as ultrafast data transmission and high-quality radar images. Nevertheless, the broad application of printed antennas at sub-THz faces a significant obstacle: surface wave (SW)-induced low radiation efficiency, which diminishes the effective radiated power of the antenna. SW diffraction at the edges of the antenna substrate also deforms the radiation pattern. As the SW intensity rises with increasing electrical thickness of the dielectric substrate [1], this factor becomes more pronounced at sub-THz frequencies compared with the microwave range.

Similar to other electromagnetic waves, SWs can exist only in the presence of a source that excites them and a medium with parameters meeting the conditions for their propagation. The grounded dielectric layer favors SW propagation [2]. SWs are usually represented in the form of a spectrum of the eigenmodes of transverse magnetic modes $TM_n$ ( $n \geq 0$ ) and transverse electric modes $TE_n$ ( $n > 0$ ). Each wave mode can propagate only when the electrical thickness of the dielectric layer and the boundary conditions at its upper and lower boundaries satisfy the propagation conditions. The $TM_0$ mode is the dominant mode because it lacks frequency restrictions. All other modes of SW have a critical frequency $f_c$ (cutoff frequency), below which they cannot propagate in a given layer. In a homogeneous dielectric layer, all modes are mutually orthogonal and do not interact. Each mode takes its share of the power from the excitation source. Hence, when designing printed antennas, the parameters of the layer are selected such that SW of higher types cannot exist in it. However, even in this case, the single dominant mode $TM_0$ can divert an unacceptably large share of the source power, thereby reducing the antenna radiation efficiency.

Three approaches for suppressing SW in printed antennas are as follows. The first approach involves devising an antenna design that either eliminates SW excitation (an ideal scenario) or minimizes the amplitude of the waves it induces. The second approach involves creating various barriers impenetrable to the SW passage. The third approach entails artificially creating boundary conditions on the upper or lower surfaces of the dielectric layer that impede SW propagation.

The first method is usually used in patch antennas [3-8] and involves purposefully disrupting the structure of the dielectric layer located between the patch and ground by creating cavities [3-5] or holes [6]. This makes it possible to reduce the local substrate permittivity, which prevents the SW antenna from exciting. The works in [7,8] provide a theoretical justification for this method and practical recommendations for its implementation.

In the second method, the SW source is enclosed by a shorting wall in the form of metal plates or a grid of short-circuiting vias [9], which limits SW localization.

The third method involves placing structures on the substrate surface that change the boundary conditions at the air–dielectric interface or damage the conductive ground integrity to change their surface impedance. This method is implemented mainly using various periodic artificial structures, metamaterial, and metasurface (MS).

A typical metastructure is an electromagnetic bandgap (EBG) [10], which may vary in shape and type [11]. These can include square lattices made of dielectrics [12] or metal rods [13], cylindrical structures [14], and mushroom-shaped structures [15-18]. Despite their proven effectiveness in suppressing SW in printed antennas [19], EBG structures are difficult to produce. A similar effect can be achieved using superstrates, but they use bulk metamaterials [20,21], thereby complicating the antenna design.

Printed antennas on a grounded single-layer dielectric are the most straightforward to manufacture. To suppress SW, parasitic elements are applied to the dielectric layer around the antenna, forming a surface impedance at the interface that is unfavorable for SW propagation [22-24]. In certain antennas [25-27], a defected ground is used for the same purposes. This results in a breach of the boundary conditions on the lower (grounded) surface of the substrate, thereby creating an unfavorable environment for SW propagation. Sometimes, when this is possible, the boundary conditions on both surfaces of the dielectric layer are changed simultaneously [28] to achieve pronounced SW suppression.

From this short review, numerous methods exist for suppressing SW in printed antennas. In most published works, SW suppression was considered an effective way to improve any parameters or characteristics of printed antennas, such as increasing radiation efficiency, increasing gain, improving matching, expanding the operating frequency band, and increasing isolation between antennas.

However, some studies failed to demonstrate that the enhanced antenna parameters result from SW suppression rather than from other electromagnetic phenomena within the studied structure. This is especially true for the study of antennas printed on short substrates where multiple reflections of SW from the edges are present. Failure to assess the SW level in such antennas raises the likelihood that the solution to the problem is facilitated by favorable interference of these waves rather than suppression.

This study considers one of the options for using an MS for effective SW suppression in a sub-THz printed antenna and provides correct numerical estimates of the level of this suppression.

## Problem formulation

Uncovering effective methods for suppressing SW in an antenna requires determining a well-posed numerical evaluation criterion for evaluating the degree of suppression. In our opinion, the most reliable for these purposes is the energy criterion, defined as follows [29]:

$$p_{sw} = \frac{P_{sw}}{P_{rad} + P_{sw}} = 1 - \frac{P_{rad}}{P_{rad} + P_{sw}} = 1 - e_{sw}, \qquad (1)$$

where $P_{rad}$ and $P_{sw}$ are powers spent by the antenna to create spatial wave and SW, respectively, $e_{sw}$ is the antenna radiation efficiency when the power $P_{rad}$ radiated into free space is useful, and $P_{sw}$ is considered as the loss power, while all other losses in the antenna, including ohmic ones, are absent:

$$e_{sw} = \frac{P_{rad}}{P_{rad} + P_{sw}}. \qquad (2)$$

However, a real antenna always contains ohmic losses; therefore, its radiation efficiency is determined as

$$e_{rad} = \frac{P_{rad}}{P_{in}} = \frac{P_{rad}}{P_{rad} + P_l}, \qquad (3)$$

where $P_{in} = P_{rad} + P_l$ is the power consumed by the antenna from the generator; $P_l = P_{sw} + P_\sigma$ is the total loss power in the antenna, which is the sum of the SW power $P_{sw}$ and the ohmic loss power $P_\sigma$.

If the ohmic losses in the substrate are small, $P_\sigma \ll P_{rad} + P_{sw}$, then (1) can be transformed into the following form:

$$e_{rad} \approx e_{sw}(1 - p_\sigma), \tag{4}$$

where $p_\sigma = P_\sigma/(P_{rad} + P_{sw}) \approx P_\sigma/P_{in}$ is the share of thermal power losses in the antenna budget.

If $p_\sigma \ll 1$, then $e_{sw} \approx e_{rad}$, we can use the antenna radiation efficiency to estimate the share of the SW power $p_{sw} \approx 1 - e_{rad}$ instead of (1). This value always gives an upper estimate of the SW level since $e_{rad}$ also considers ohmic losses even if they are small. When antenna studies are conducted using a full-wave numerical simulation, the value $\delta_{sw}$ can be refined by setting all losses in the antenna to zero. The difference between the $e_{sw}$ and $e_{rad}$ of the antenna with losses obtained in this case gives the error value $\delta_{sw}$ when using the approximation $e_{sw} \approx e_{rad}$.

We will study SW properties and methods for their suppression in printed antennas using a simple, well-studied antenna—a symmetrical linear strip dipole printed on a flat layer of a homogeneous grounded dielectric of unlimited length and width. This choice guarantees the presence of only traveling waves generated by the antenna in the dielectric and free space and will also allow the results obtained to be compared with those of other studies. Fig. 1 shows the dipole under study in Cartesian and spherical coordinate systems.

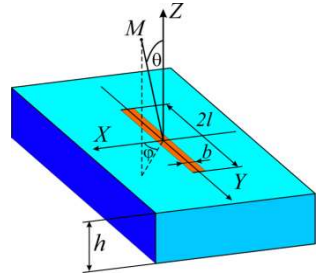

**Figure 1.** A single microstrip dipole.

We study the parameters of this dipole in the frequency range 100–116 GHz, assuming that the substrate is a dielectric, such as gallium nitride, with relative permittivity $\varepsilon_r = 9.4$ and permeability $\mu_r = 1$, respectively, and dielectric loss tangent $\tan\delta = 0.0005$. The thickness of the substrate is $h = 0.2$ mm, and the dipole width and length are $b = 0.05$ mm and $2l = 0.45$ mm, respectively, which ensure its resonance at a frequency of 108 GHz.

To determine the SW mode spectrum that can propagate in this substrate, the following expressions for calculating cut frequencies are used [1]:

– for $TM_n$ modes

$$f_{cn} = \frac{2nc}{4h\sqrt{\varepsilon_r - 1}}, \quad n = 0, 1, 2, \ldots; \tag{5}$$

– for $TE_n$ modes

$$f_{cn} = \frac{(2n-1)c}{4h\sqrt{\varepsilon_r - 1}}, \quad n = 1, 2, 3, \ldots, \tag{6}$$

where $c$ is the light velocity in vacuum.

Substituting the values of n into equations (5) and (6), we are convinced that only one $TM_0$-mode exists in the frequency range under consideration, since the nearest $TE_1$-mode cutoff frequency is 129.3 GHz.

The SW wave number can be obtained from the dispersion equation [1]

$$\varepsilon_r\sqrt{k_t^2 - k_0^2} - \sqrt{k_2^2 - k_t^2}\ \tan\left(h\sqrt{k_2^2 - k_t^2}\right) = 0, \qquad (7)$$

where $k_0 = 2\pi/\lambda_0$ is the wave number of free space, $k_2 = k_0\sqrt{\varepsilon_r} = \sqrt{k_t^2 + k_z^2}$ is the wave number of the dielectric, $k_t$ is the SW wave number along the interface, and $k_z$ is the transverse wave number in the dielectric layer.

Using Equation (7), we calculated the wave numbers $k_t$ and $k_z$ for the structure under study (Fig. 1). The relationships between the wavelengths $\lambda_0$, $\lambda_2$, $\lambda_t$, and $\lambda_z$ corresponding to the abovementioned wave numbers can be seen in Fig. 2, which shows their frequency dependence at 80–160 GHz.

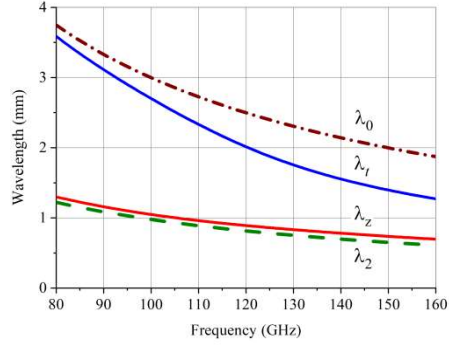

**Figure 2.** Frequency dependence of wavelengths in the grounded dielectric layer and free space.

According to preliminary estimates, the radiation efficiency of such a dipole on a lossless substrate is approximately 30%–35% [30] at 80–160 GHz. This means that most of the source power is spent creating the SW, which significantly worsens the antenna power parameters. Below, we show how to markedly increase the radiation efficiency of a dipole using our proposed MS cell to suppress SW in this structure. The experiment was numerically conducted using Altair Feko 2022 software [31], which provides full-wave simulation using the method of moments.

## MS Cell

As known [1], a necessary condition for the propagation of SW $TM_0$ on a grounded dielectric layer is the inductive nature of its surface impedance. Preventing the propagation of these waves requires changing the nature of the surface impedance to a capacitive one. This can be achieved by placing an MS on the top side of the layer, which conducts the necessary impedance transformation. Therefore, we developed an MS element called quad-split rings (QSR), as shown in Fig.3. It is a set of four-square rings with gaps shifted to the corners, which are concentrated in the center of the structure. Despite the asymmetry of each ring, the entire element has double symmetry.

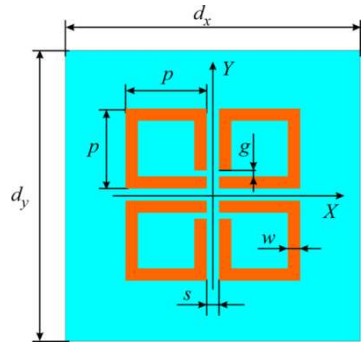

**Figure 3.** The layout of the proposed QSR unit cell, $p = 165$ μm, $s = w = 25$ μm, $g = w/2$.

Using the transverse resonance method, we studied the surface impedance of the dielectric layer (Fig. 1) covered with MS with unit cell dimensions $d_x \times d_y$ [32, 33]. This method makes it possible to determine the surface resistance of SW propagating in this layer from the known reflection coefficient $R$ of a spatial plane electromagnetic wave incident vertically on this layer.

$$Z_S = z_S = Z_0/y_S ; \tag{8}$$

$$y_S = \frac{1-R}{1+R} + j\sqrt{\varepsilon_r}\cot(k_2 d) - j\frac{k_0}{k_z}\varepsilon_r \cot(k_z d), \tag{9}$$

where $z_S$ and $y_S$ are the normalized surface impedance and admittance, respectively, of the grounded dielectric layer for the $TM_0$-mode.

We determined the surface impedance of an infinite periodic QSR grid lying on a grounded dielectric layer (Fig. 1) for which we calculated its reflection coefficient when a plane electromagnetic wave of linear polarization with electric field strength $\vec{E}^i = \vec{y}^0 E_0 e^{jk_0 z}$ was incident. Fig. 4 shows the frequency dependence of the reflection coefficients $R_{0.5}$ and $R_{0.6}$ of MS with QSR cell sizes $d_x = d_y = 0.5$ mm and $d_x = d_y = 0.6$ mm, respectively, obtained by simulation at 80–160 GHz.

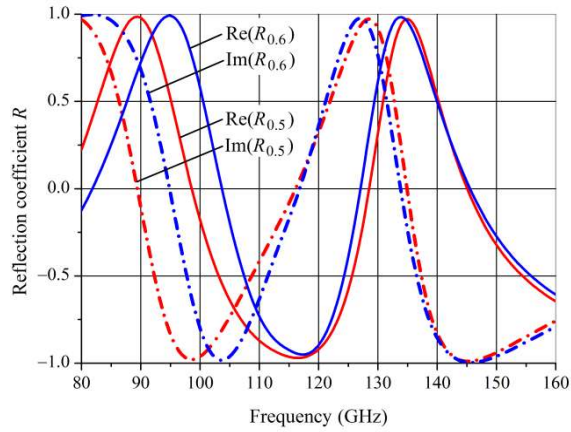

**Figure 4.** Reflection coefficients R05 and R06 for QSR cell dimensions of $d = 0.5$ mm and $d = 0.6$ mm, respectively.

Using the reflection coefficients found and the previously obtained wave numbers $k_t$ and $k_z$ (7), the surface impedance $z_S$ (8) of the MS versions under study was calculated. Fig. 5(a) shows the frequency dependence of the imaginary part of the normalized surface reactance $x_S$ for both versions of the MS: 1 is the solid line and 2 is the dashed line. In addition, Fig. 5(b) shows the same in detail in a narrow frequency band of 100–120 GHz. Fig. 5 shows a dash-dotted line that indicates a similar surface reactance of a layer without an MS. Fig. 5 also shows that resistances 1 and 2 have one resonance and two antiresonances at 80–160 GHz.

The frequencies of the upper antiresonances 1 and 2 are almost the same. The frequencies of resonances 1 and 2 are very close to each other, with the second lying slightly higher in frequency $f_1 = 116.11$ GHz and $f_2 = 116.78$. The frequencies of the lower antiresonances markedly differ (93.25 GHz and 98.75 GHz), and the difference between them is 5.5 GHz. Thus, with an increase in the period $d$ of the MS, the frequency band between its antiresonances narrowed; in this case, this band decreased from 47 GHz to 41.5 GHz with a decrease in $d$ by 0.1 mm. Fig. 5(b) shows that in the frequency band of interest, 100–116 GHz, the imaginary part of the surface impedance $x_S$ of both MS versions for the SW is negative, unlike the $x_S$ dielectric layer without MS (upper curve). Hence, the imposition of MS on the dielectric layer surface changes its positive surface reactance to negative, which should prevent SW propagation, thereby suppressing them in printed antennas.

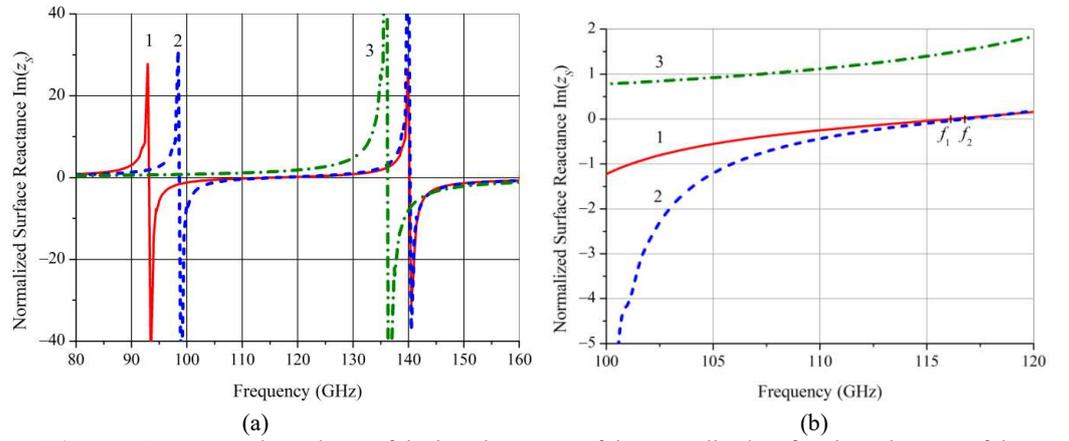

**Figure 5.** Frequency dependence of the imaginary part of the normalized surface impedance $z_S$ of the grounded layer with and without MS: 1, MS with unit cell 0.5 mm; 2, MS with unit cell 0.6 mm; 3, no MS; (a) general view and (b) detailed view.

## SW suppression in a printed dipole antenna

We explored the possibility of using MS to suppress SW in a printed dipole antenna (Fig. 1). To achieve this, we formed two identical QSR arrays $M \times N$ arranged symmetrically relative to the dipole center (Fig. 6(a)). In general, each array is two-dimensional with a period $d_x \times d_y$, and the first row of the array is located at a distance $d_0$ from the $X$ axis. This MS should become an obstacle to the $TM_0$-mode, the maximum propagation of which is directed along the $Y$ axis [34].

We initiated the analysis with the simplest option when the MS array degenerated into one QSR element $M \times N = 1 \times 1$ (Fig. 6(b)). Having performed a full-wave electromagnetic simulation of this structure in the Feko environment, we determined the parameters and characteristics of the dipole required for analysis in the presence of two QSRs, namely, radiation pattern, gain, directivity, and radiation efficiency.

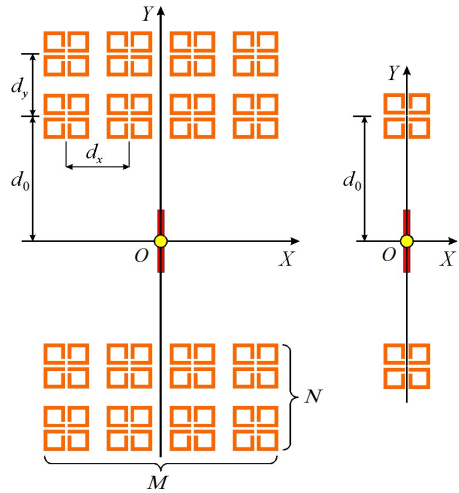

**Figure 6.** Dipole with two QSR arrays of dimension: (a) $M \times N$; (b) $1 \times 1$.

Fig. 7 shows the frequency dependence of the dipole radiation efficiency for different distances $d_0$. It follows from these results that QSR markedly affects the dipole radiation efficiency by increasing or decreasing it. The highest efficiency is achieved at 108–110 GHz for $d_0$ of 1.2–1.4 mm (the distance $d_0 = 1.2$ mm coincides with half the SW length at 108 GHz). Reducing or increasing the distance $d_0$ beyond these limits decreases the achievable dipole radiation efficiency. Thus, considering that the radiation efficiency of a single dipole is 31.2%, the use of only two QSRs makes it possible to increase it by almost 2.5 times.

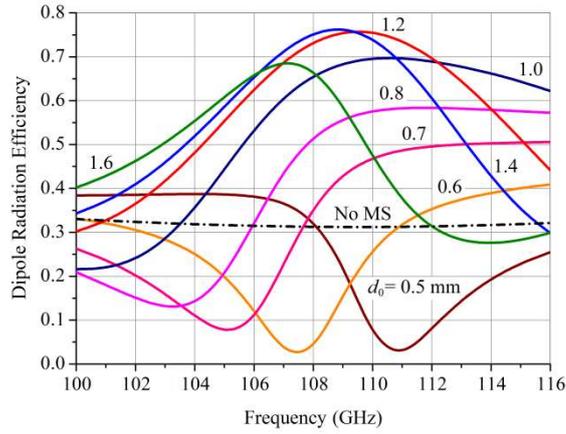

**Figure 7.** Frequency dependence of dipole radiation efficiency with and without QSR for various $d_0$ values.

To separate the losses associated with the finite conductivity of the dielectric, we repeated the simulation of this structure, setting $\tan\delta = 0$. We found that the losses in the dielectric (6) are negligible ($p_\sigma \approx 0.74\%$), and they were disregarded when estimating the share of power $p_{sw} \approx 1 - e_{rad}$ spent by the source to create the SW. Comparing the $p_{sw}$ of the dipoles with and without MS, we found that the use of these two QSRs facilitated the suppression of the SW power by approximately 2.9 times or 4.6 dB.

Adding a QSR to a dipole affects not only its radiation efficiency but also its other characteristics. Fig. 8 shows the normalized radiation patterns of the dipole with two QSRs at different frequencies for $d_0 = 1.2$ mm. The dipole radiation pattern (RP) with MS at 100 GHz remains the same as that without MS, as well as their radiation efficiencies at 100 GHz (Fig. 7), suggesting that, at this frequency, MS does not affect the dipole parameters. However, with an increase in frequency from 100 GHz to 116 GHz, this impact increases, which manifests itself in a narrowing of the dipole RP in the $E$-plane $2\theta_{0.5}^E$ from 100° up to 46° (Fig.8(a)), and the expansion of its RP in the $H$-plane $2\theta_{0.5}^H$ from 101° to 118° (Fig. 8(b)).

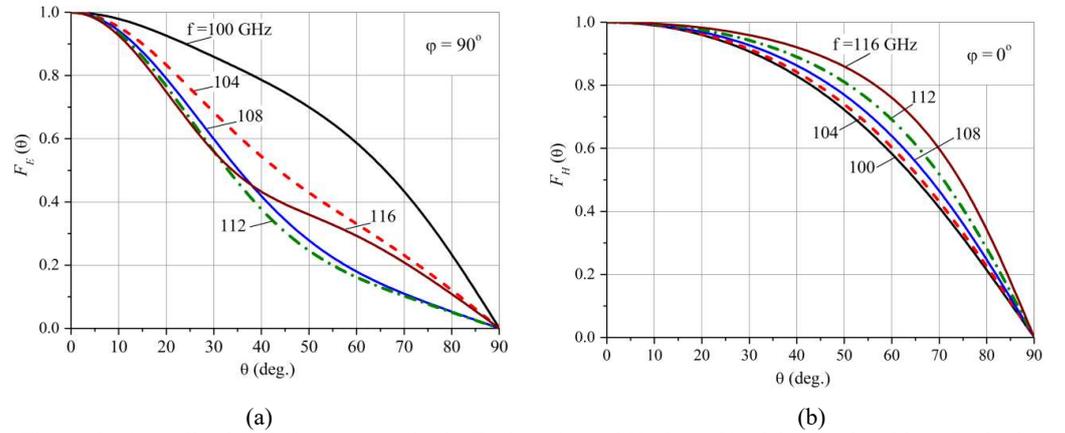

(a)      (b)

**Figure 8.** Normalized radiation patterns in the E-plane (a) and H-plane (b) of the dipole with QSRs in the frequency band 100–116 GHz ($d_0 = 1.2$ mm)

Fig. 9 illustrates the differences in dipole RPs with and without MS at 108 GHz. This variation in the RP of a dipole with MS affects its other parameters, particularly its directivity and gain. Fig. 10 shows their frequency dependence, which peaks near 110 GHz, while the same parameters of the dipole without MS remain practically constant. The greatest increase due to MS was the dipole gain (7.6 dB), to which both radiation efficiency (5.6 dB) and directivity (2 dB) contributed.

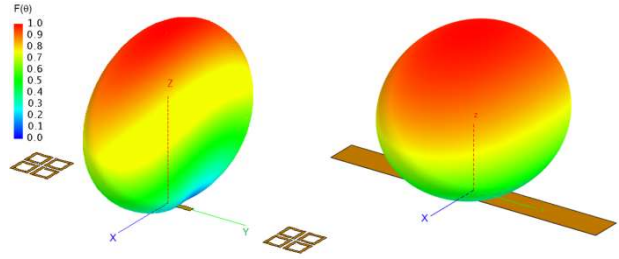

**Figure 9.** 3D radiation patterns of the dipole with (a) and without (b) QSRs at 108 GHz

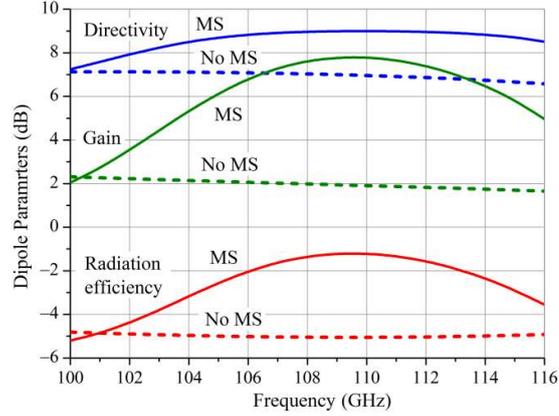

**Figure 10.** Directivity, gain, and radiation efficiency of dipoles with and without MS

Now, we consider the effect of increasing the number of elements in the MS on dipole efficiency. Fig. 11(a) shows the frequency dependence of the efficiency of a dipole with four MS $N_x \times N_y$ options, one of which was already discussed above $1 \times 1$, and three new ones, $2 \times 1$, $4 \times 1$ and $4 \times 2$. Distance $d_0 = 1.2$ mm and period $d = d_x = d_y = 0.6$ mm were assumed in the calculations.

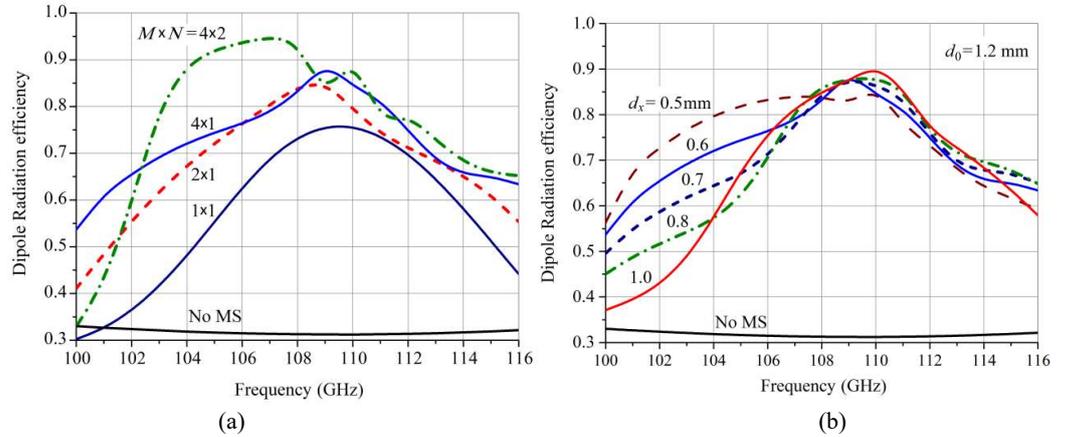

**Figure 11.** Frequency dependence of the radiation efficiency of the dipole: a) with MS of various structures at fixed $d_0$ = 1.2 mm and $d$ = 0.6 mm; b) with 4 × 1 MS at fixed $d_0$ = 1.2 mm and various $d$

These graphs show that increasing the number of elements $N_x$ in the MS to four can yield a dipole efficiency of 87% (i.e., approximately 11%–12% more than in the case of $1 \times 1$ MS). Calculations showed that a further increase in the number of elements $N_x$ in a single-row MS only slightly increased the dipole efficiency; therefore, we added one more row to it. The dashed-dotted curve in Fig. 11(a) shows the frequency dependence of the dipole efficiency with a $4 \times 2$ MS, which peaks (93.6%) at 107 GHz, which is 6% higher than the maximum with a single-row MS.

Until now, we have used MS with a fixed distance between the elements in the row $d_x = 0.6$ mm, so we should consider the behavior of the dipole efficiency when varying the distance $d_x$. Fig. 11(b) shows that with $d_x$ increasing from 0.6 mm to 1 mm, the achievable value of the maximum radiation efficiency of the dipole with $4 \times 2$ MS increases slightly from 87.6% to 89.5%; however, there is a tendency to narrow the frequency band, in which a fixed efficiency value is maintained.

The analysis results showed that the use of MS triggered an increase in the dipole radiation efficiency by 3 times from 31.2% to 93.5% due to SW suppression by 10.3 dB. Fig. 12 shows visual evidence of SW suppression in the dipole antenna. Here, we show several distributions of the tangential component of the Poynting vector modulus of the SW created by a dipole on the substrate surface in the presence of different $M \times N$ MSs. All distributions are calculated assuming the same power ($P_{in} = 15$ mW) delivered to the dipole from the source, and each of them refers to the frequency at which the maximum radiation efficiency of the corresponding structure is observed (Fig. 11(a)).

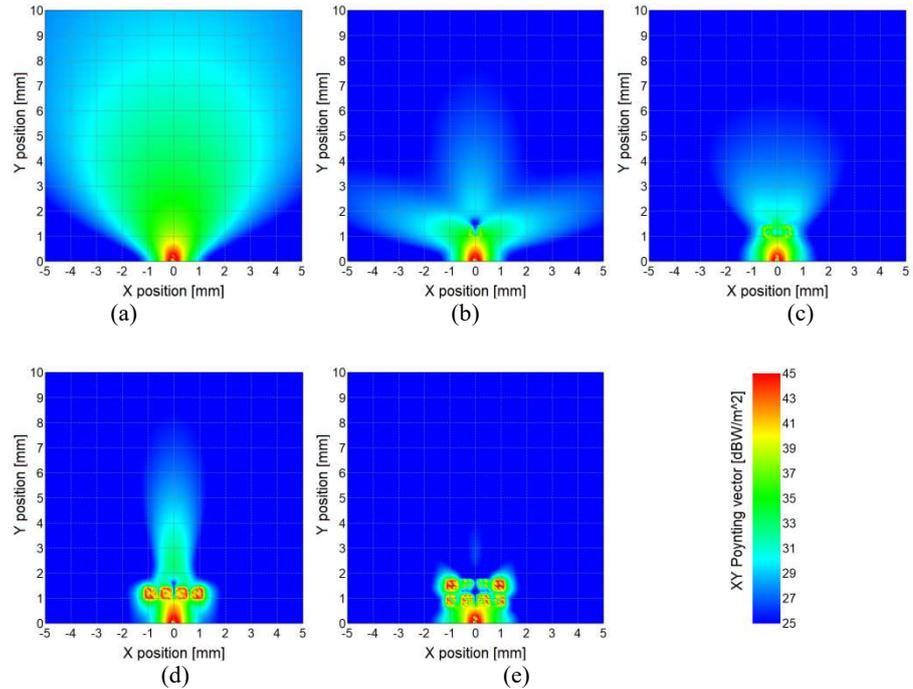

**Figure 12.** SW Poynting vector modulus distribution in the dipole antenna with different MSs: (a) no MS, 108 GHz; (b) 1 × 1 MS, 108 GHz; (c) 2 × 1 MS, 108 GHz; (d) 4 × 1 MS, 109 GHz; (e) 4 × 2 MS, 107 GHz.

These pictures show that the torch-shaped flow of SW power created by a single dipole in the dielectric layer gradually fades as an increased number of elements are added to the MS.

## SW suppression in a printed dipole array

Inspired by the results of SW suppression in a single dipole printed antenna, we decided to apply QSR structures to a printed antenna array. For this study, we chose a four-element series-fed dipole array (Fig. 13) printed on the same substrate as the dipole (Fig. 1). The symmetrical dipoles of length $2l$ and width $b$ are fed by a coupled microstrip line comprising two coplanar strips of width $s$ separated by a gap of width *gap*.

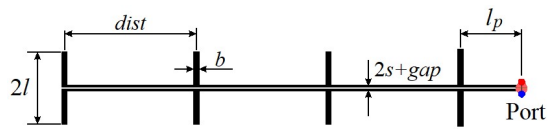

**Figure 13.** Four-element serial-fed dipole array, $2l$ = 0.676 mm, $b$ = 0.05 mm, $s$ = 0.02 mm, *gap* = 0.01 mm, *dist* = 1.2 mm, $l_p$ = 0.535 mm.

At 106–112 GHz, the array forms the main beam oriented along the normal to its plane. The array's radiation pattern in the H-plane is shown in Fig. 14(a), and that in the E-plane is shown in Fig. 14(b). It has one main beam, which narrows slightly in both planes with increasing frequency, from $2\theta_{0.5}^H = 33.5°$ and $2\theta_{0.5}^E = 91°$ at 106 GHz to $2\theta_{0.5}^H = 30.5°$ and $2\theta_{0.5}^E = 86°$ at 112 GHz, thereby increasing the maximum array gain from 7 dB to 7.8 dB. The slight beam squinting (Fig. 14) inherent in most series-fed arrays reduces the array gain in the normal direction within the frequency band insignificantly (< 0.6 dB). The array's efficiency in the considered frequency band is approximately 40%. Despite good parameters for its class, low radiation efficiency notably diminishes its energy potential. We now explore how to enhance the array's efficiency and improve its parameters using the proposed MS.

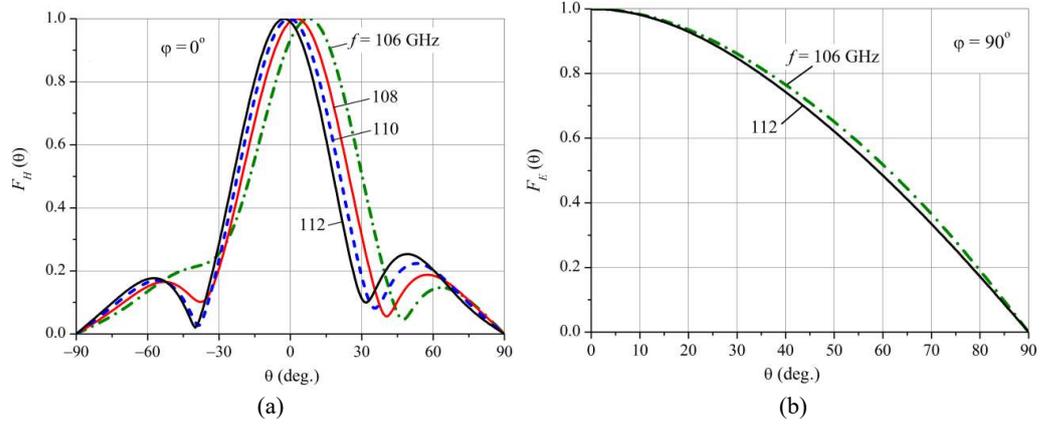

**Figure 14.** Normalized radiation patterns of the dipole array: (a) in the H-plane and (b) in the E-plane.

For consideration, we selected five options for modifying the dipole array, shown in Fig. 15, which differ in the number of QSR elements in the MS: 3×1 (Fig. 15(a)), 4×1 (Fig. 15(b)), 5×1 (Fig. 15(c)), 7×1 (Fig. 15(d)), 8×1 (Fig. 15(e)), 9×1 (Fig. 15(f)), 9×2 (Fig. 15(g)), and 10×2 (Fig. 15(h)). The MS grating period $d$ along both coordinate axes was chosen as a multiple of $dist/2$.

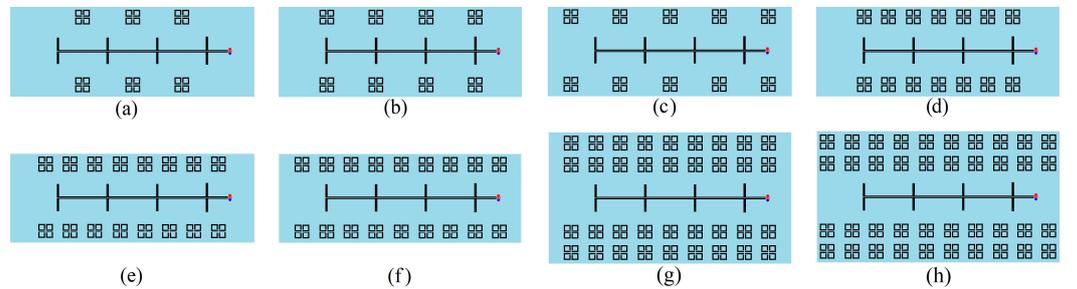

**Figure 15.** The dipole array with different MS of M × N: (a) 3 × 1; (b) 4 × 1; (c) 5 × 1; (d) 7 × 1; (e) 8 × 1; (f) 9 × 1; (g) 9 × 2; (h) 10 × 2.

The frequency dependence of the efficiency of the modified dipole arrays (Fig. 15), as well as the original array without MS (Fig. 13), obtained from the simulation, is shown in Fig. 16. As the single-row MS length increased, the achievable efficiency of the dipole array increased. Here, almost identical maximum efficiencies of 84.6% and 85.6% resulted from dipole arrays with 5×1 и and 9×1 MS, respectively, where the MSs protrude beyond the edges of the dipole array by a distance $d$. As calculations have shown, a further increase in the single-row MS length does not markedly increase the efficiency of this array. However, this efficiency can be increased even more by adding another row to the MS.

Fig. 16 shows the radiation efficiency of the dipole array with a double-row 9×2 MS, which reaches an extremely high value of 96.5%. If we consider that part of the input power of the dipole array is taken away by thermal losses $p_\sigma \approx 0.8\%$, the portion of it that is spent on creating SW can easily be calculated $p_{sw} = 1 - p_\sigma - e_{rad} = 2.7\%$, which is 13.5 dB lower than that in the dipole array without MS ($p_{sw} \approx 59\%$).

Fig. 16 also shows that expanding the MS to $10\times 2$ MS practically does not increase the maximum radiation efficiency of the dipole.

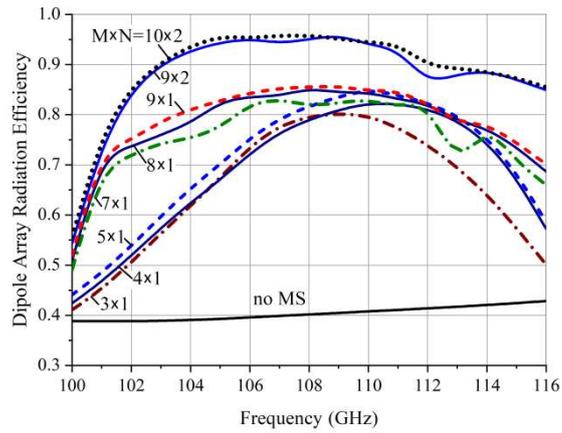

**Figure 16.** Frequency dependence of the radiation efficiency of dipole arrays with different MS

The images depicted in Fig. 17 enable observation of the variations in the distribution of the modulus of the Poynting vector SW in an antenna array with a constant input power (15 mW), correlating with the increase in the number of MS elements. Each presented power flow density distribution was calculated at the frequency at which the radiation efficiency of the related structure was highest. These images provide visual evidence of SW suppression in a particular antenna array version.

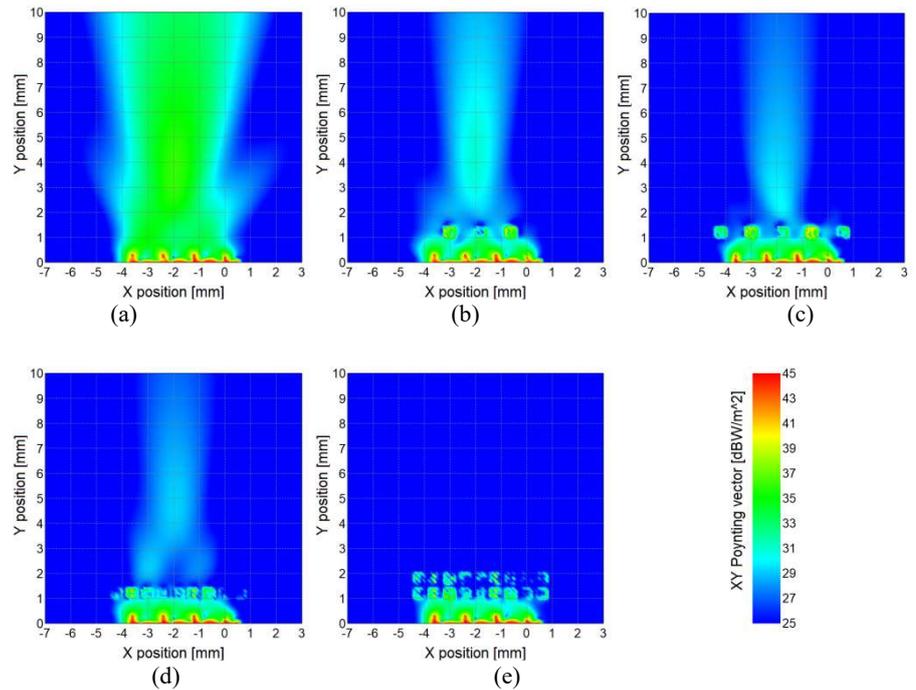

**Figure 17.** SW Poynting vector modulus distribution in the dipole array with different MSs: (a) no MS, 108 GHz; (b) 3 × 1 MS, 108 GHz; (c) 5 × 1 MS, 108 GHz; (d) 9 × 1 MS, 109 GHz; (e) 9 × 2 MS, 108 GHz.

Now, we briefly examine the influence of QSR MS on the array radiation pattern. Comparing the RP of the dipole array with MS (Fig. 18) with a similar RP without MS (Fig. 14), we can draw the following conclusions. The RP in the *H*-plane of both arrays differs slightly, but in the *E*-plane, the differences are obvious. Adding MS to the dipole array not only increases the frequency dependence of its RP but also changes the behavior of this dependence.

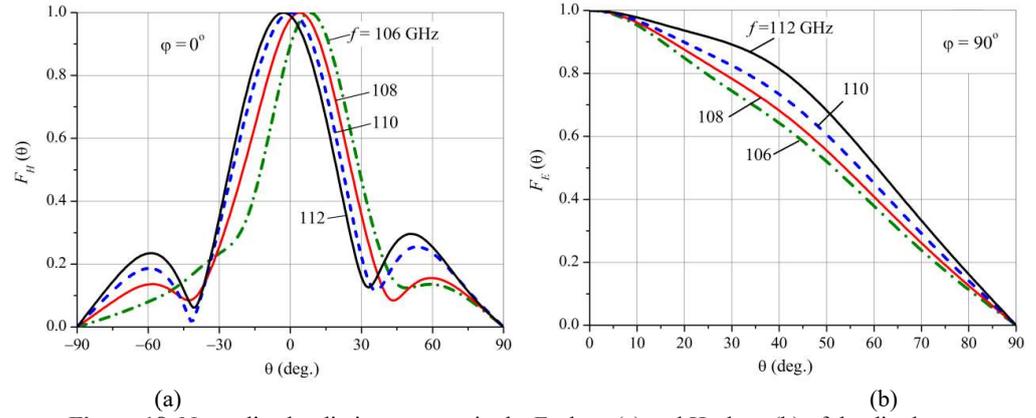

**Figure 18.** Normalized radiation patterns in the E-plane (a) and H-plane (b) of the dipole array with 9 × 2 MS at 106–112 GHz.

The narrowing of the beam in a no MS array with increasing frequency is a normal phenomenon that can be easily explained by increasing the electrical dimensions of an antenna. However, in the dipole array with MS, we observe the opposite process: its main beam expands with increasing frequency, and this expansion is several times greater than the narrowing of the beam of the array without MS. This anomalous phenomenon results from the fact that a dipole array with an MS is a complex structure in which both the dipoles and the elements of the MS participate in the radiation of spatial waves. Therefore, the width of its RP in the $E$-plane $2\theta_{0.5}^E$ depends on the currents induced on the MS elements: the closer their amplitudes and phases are to those on the dipoles, the larger the equivalent size of the antenna in the $E$-plane and the narrower its beam. This is exactly what happens in this dipole array; therefore, its beam narrows in the $E$-plane with decreasing frequency.

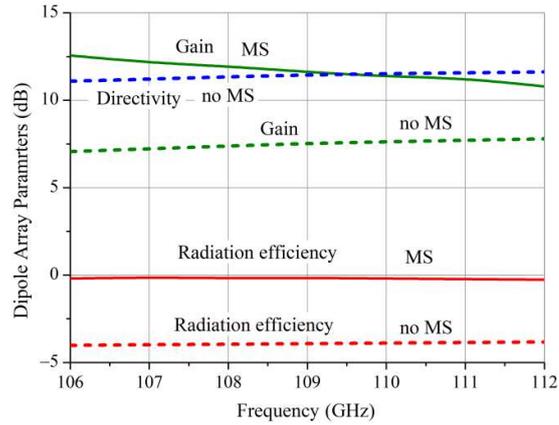

**Figure 19.** Directivity, gain, and radiation efficiency of dipole arrays with and without MS

In summary, the frequency dependence results of the dipole array parameters with and without MS, shown in Fig. 19, confirm the effectiveness of using MS QSR to increase its radiation efficiency and gain.

## Conclusion

The possibility of increasing the radiation efficiency of printed dipoles and dipole arrays by suppressing SW using MSs was studied. As an element of the MS, a QSR with gaps shifted to the structure center was proposed. The properties of this MS were investigated using the transverse resonance method, which revealed that MS creates a negative surface reactance that prevents SW propagation in the grounded dielectric layer.

The effectiveness of the developed MS was tested on two antennas: a linear symmetric strip dipole and a 4-element serial-fed dipole array, both printed on a low-loss dielectric layer of 0.2 mm thickness. The studies were conducted in the 110 GHz range, where the substrate was electrically thick, which is why most of the power delivered to the antenna was spent creating the SW.

The dipole antenna used MS in the form of two small QSR arrays symmetrically placed near the dipole arms. Arrays comprising just one QSR each have a substantial impact on the SW level, which enables control of the SW level by adjusting the distance to the dipole. When the distance was close to half the SW length, the radiation efficiency of the dipole peaked (~75%), which corresponded to SW suppression of 4.6 dB. Increasing the number of QSR elements in the MS could yield marked SW suppression, resulting in greater dipole radiation efficiency. A dipole radiation efficiency of 93.5% or SW suppression of 10.3 dB was achieved when the dimension of the QSR arrays was increased to 4 × 2.

The proposed MS effectively suppressed SW (up to 13.5 dB) in the dipole array, thereby increasing the radiation efficiency from 41% to 96.5%. The presented images of the SW power flow density distribution confirm that the increased radiation efficiency of the antennas results from SW suppression.

**Author contributions**
Y.M.Y.: Conceptualization, funding acquisition, project administration; supervision, writing—review and editing.
P.L.T.: Methodology, software, investigation, data processing, formal analysis, visualization, validation, and writing—original draft.

**Funding**
This work was supported with basic funding from the Institute of Radioelectronics and Multimedia Technology, Warsaw University of Technology.

**Competing interests**
The authors declare no competing interests.

**Additional information**
Correspondence and requests for materials should be addressed to Y.M.Y.